\documentclass[preprint,aps,prd,showpacs,superscriptaddress,nofootinbib,tightenlines]{revtex4}

\usepackage{amsfonts}
\usepackage{multirow}
\usepackage{mathrsfs}
\usepackage{graphicx}
\usepackage{amsmath}
\usepackage{amssymb}
\usepackage{bm}
\usepackage{bbm}
\usepackage{color}
\usepackage{hyperref}
\usepackage{diagbox}
\usepackage{slashed}
\usepackage{xparse}

\def\OMIT#1{}

\newcommand{\beq}{\begin{equation}}
\newcommand{\eeq}{\end{equation}}
\newcommand{\bqa}{\begin{eqnarray}}
\newcommand{\eqa}{\end{eqnarray}}

\newcommand{\bseq}{\begin{subequations}}
\newcommand{\eseq}{\end{subequations}}

\begin{document}

\title{Electromagnetic and hadronic decay of fully heavy tetraquark}


\author{Wen-Long Sang~\footnote{wlsang@swu.edu.cn}}
 \affiliation{School of Physical Science and Technology, Southwest University, Chongqing 400700, China\vspace{0.2cm}}

\author{Tao Wang}
 \affiliation{School of Physical Science and Technology, Southwest University, Chongqing 400700, China\vspace{0.2cm}}

 \author{Yu-Dong Zhang~\footnote{ydzhang@mails.ccnu.edu.cn}}
 \affiliation{
 	Institute of Particle Physics and Key Laboratory of Quark and Lepton Physics (MOE),
 	Central China Normal University,Wuhan, Hubei 430079, China\vspace{0.2cm}}

\author{Feng Feng\footnote{f.feng@outlook.com}}
\affiliation{China University of Mining and Technology, Beijing 100083, China\vspace{0.2cm}}
\affiliation{Institute of High Energy Physics, Chinese Academy of
Sciences, Beijing 100049, China\vspace{0.2cm}}

\date{\today}

\begin{abstract}

In this study, we compute the electromagnetic and hadronic decay widths of the S-wave fully heavy tetraquark $T_{4Q}$ ($Q=c$ or $b$) at 
lowest order in $\alpha_s$ and $v$,
in the framework of nonrelativistic QCD. 
The short-distance coefficients are determined through the standard procedure of matching.
The nonperturbative long-distance matrix elements are related to the phenomenological four-body Schr\"odinger wave functions at the origin, whose values are taken
from literature.  
The branching fractions are predicted to be around $10^{-4}$ and  $10^{-7}-10^{-6}$ for the $T_{4c}$ hadronic decay and electromagnetic decay, respectively.
Combing our results with the $T_{4c}$ production cross sections at the LHC, we also predict the event numbers for various decay channels.
With integrated luminosity $\mathcal{L}=100 \,{\rm fb}^{-1}$,  it is expected that the event numbers can reach $10^3$ for $T_{4c}\to \gamma\gamma$,
and $10^6$ for $T_{4c}\to {\rm LH}$, at the LHC.   The detecting prospect is promising.
In addition, the decay widths of $T_{4b}$ are estimated based on simple dimensional analysis as well as velocity scaling rule. 
\end{abstract}

\maketitle

\section{Introduction}
Searching for exotic states is full of challenges and opportunities. 
Compared with conventional hadrons, exotic states may provide a better environment  to reveal the nonperturbative nature of QCD. 
In 2020, LHCb collaboration reported a narrow resonance 
around 6.9 GeV with the significance larger than $5\sigma$, dubbed as $X(6900)$,  a broad structure with the mass ranging from 6.2 GeV to 6.8 GeV,
and a hint for another structure around 7.2 GeV  in the di-$J/\psi$ invariant mass spectrum~\cite{LHCb:2020bwg}.  The narrow structure contains
at least $cc\bar{c}\bar{c}$ configuration and therefore is naturally considered as a fully charmed tetraquark. 
By assuming that the nonresonant single-parton scattering (NRSPS) continuum is not disturbed, the mass and width of $X(6900)$ 
are determined to be $M=6905\pm 11\pm 7\,{\rm MeV}$ and $\Gamma=80\pm 19 \pm 33 \,{\rm MeV}$, respectively, 
while, they become
$M=6886\pm 11\pm 11\,{\rm MeV}$ and $\Gamma=168\pm 33 \pm 69 \,{\rm MeV}$ when assuming the NRSPS continuum interferes with a broad structure above the di-$J/\psi$ mass threshold.

Soon afterward the $X(6900)$ was confirmed in the di-$J/\psi$ channel by the CMS collaboration~\cite{Zhang:2022toq,CMS:new1,CMS:2023owd}, and the di-$J/\psi$ 
as well as $J/\psi \psi(2S)$ channels by the ATLAS collaboration~\cite{Xu:2022rnl}, respectively.  
For CMS, the mass and width are determined to be $M=6927\pm 9\pm 4\,{\rm MeV}$, $\Gamma=122^{+24}_{-21} \pm 18 \,{\rm MeV}$, or 
$M=6847^{+44+48}_{-28-20}\,{\rm MeV}$ and $\Gamma=191^{+66+25}_{-49-17} \,{\rm MeV}$ when considering or not considering the interferences among resonances, respectively~\cite{CMS:2023owd}.  
For ATLAS, the mass and decay width are $6.87 \pm 0.03 ^{+0.06}_{-0.01}\,{\rm GeV}$ and $0.12 \pm 0.04 ^{+0.03}_{-0.01} \,{\rm GeV}$, respectively ~\cite{Xu:2022rnl}.
Moreover, some new resonances are observed, i.e., $X(6600)$ and $X(7200)$ by CMS~\cite{Zhang:2022toq,CMS:new1,CMS:2023owd}, and $X(6200)$, $X(6600)$ and $X(7200)$ by ATLAS~\cite{ATLAS:new1}.  
On the other hand, in bottom quark sector, an observation of $\Upsilon$ pair was reported
by the CMS Collaboration~\cite{CMS:2016liw}, however the structure was not yet confirmed by the later analysis from CMS~\cite{CMS:2020qwa}. In addition, the LHCb collaboration studied
the $\Upsilon\mu^+\mu^-$ invariant mass spectrum, however none of reliable signals were observed~\cite{LHCb:2018uwm}. 

On the theoretical side, the explorations on fully heavy tetraquark date back to 1970s~\cite{Iwasaki:1976cn,Chao:1980dv,Ader:1981db}. 
The discovery of $X(6900)$ in experiment has inspired extensive theoretical investigations.
Various approaches are adopted to understand the nature of these fully charmed tetraquarks, e.g.,  constituent quark models~\cite{Badalian:1985es,Berezhnoy:2011xn,Wu:2016vtq,Barnea:2006sd,Liu:2019zuc,Wang:2019rdo,Bedolla:2019zwg,Deng:2020iqw,Jin:2020jfc,Lu:2020cns,Zhang:2022qtp},
QCD sum rules~\cite{Wang:2017jtz,Chen:2016jxd,Wang:2020ols,Wu:2022qwd},  diffusion Monte Carlo~\cite{Bai:2016int,Gordillo:2020sgc}, Lattice~\cite{Hughes:2017xie},  partial wave analysis~\cite{Liang:2021fzr}, color evaporation model~\cite{Carvalho:2015nqf,Maciula:2020wri}.   
Albeit many theoretical researches on these fully charmed tetraquark states are performed, 
the interpretations for them are still controversial. 
Many interpretations exist in literature,   such as P-wave tetraquark~\cite{Chen:2020xwe,liu:2020eha},   radial excitation~\cite{Giron:2020wpx,Wang:2020ols,Karliner:2020dta},
$\bar{\chi}_{c0}\chi_{c0}$ molecular state~\cite{Albuquerque:2020hio}, hybrid state~\cite{Wan:2020fsk}, 
resonance formed in charmonium-charmonium scattering or kinematic cusp arising from final-state interaction ~\cite{Dong:2020nwy,Guo:2020pvt}. 
For more references, we refer the reader to Ref.~\cite{Chen:2022asf} and references therein.

While the spectra have been widely studied, the decay property and production mechanism of the fully-heavy tetraquarks are relatively less investigated,
particularly based on model independent methods.
In Ref.~\cite{Feng:2020riv}, a factorization formula for  S-wave $T_{4c}$ inclusive production is presented in the framework of nonrelativistic QCD (NRQCD)~\cite{Bodwin:1994jh}
(the similar idea can be also found in~\cite{Ma:2020kwb,Zhu:2020xni}).
A key observation is that, prior to hadronization, two charm quarks and two anticharm quarks have to be created at a rather short distance $< 1/m_c$, thus
one can invoke asymptotic freedom to factorize the production rate as the product of the perturbative calculable short-distance part and
the nonperturbative long-distance part.  By employing this factorization formula, the $T_{4c}$ production at the LHC was studied in Refs.~\cite{Feng:2020riv,Feng:2023agq},  and production at the B factory was investigated in Refs.~\cite{Feng:2020qee,Huang:2021vtb}.  
Analogous to the case in quarkonium sector,  one anticipates that the NRQCD factorization ansatz should  hold for the 
fully heavy tetraquark inclusive and electromagnetic decay.  
This work aims to compute the decay widths for $T_{4Q}\to \gamma\gamma$ and $T_{4Q}\to {\rm LH}$, where ${\rm LH}$ denotes the light hadrons. 
Note that, by assigning $T_{4Q}$ to be an S-wave tetraquark,  the feasible $J^{PC}$ quantum number of $T_{4Q}$ can be
$0^{++}$, $1^{+-}$ and $2^{++}$. 

The paper is organized as follows. In Sec.~\ref{sec-framework}, we present the factorization formulas for $T_{4Q}$ hadronic and electromagnetic decay, and
sketch the procedure to match the short-distance coefficient (SDC). 
In Sec.~\ref{sec-sdcs},  we describe the computing techniques, and present the analytic expressions for the SDCs. 
Sec.~\ref{sec-phen} is devoted to phenomenological predictions and discussions.  We make a summary in Sec.~\ref{sec-sum}.

\section{Theoretical framework~\label{sec-framework}}

\subsection{NRQCD factorization for $T_{4Q}$ decay}
In Ref.~\cite{Feng:2020riv}, the NRQCD factorization formula for $T_{4Q}$ inclusive production was proposed.
It is straightforward to convert the factorization formula into that for the $T_{4Q}$ electromagnetic and hadronic decay.
Specifically, we can express the decay widths of $T_{4Q}\to {\rm LH}$ at the lowest order in velocity as follows~\footnote{Due to the Landau-Yang theorem,  the $T_{4Q}^{1^{+-}}$ into double photons or double gluons is strictly forbidden.  Moreover, $T_{4Q}^{0^{++}}$ or $T_{4Q}^{1^{+-}}$ decay into a pair of light quarks is also forbidden according to the helicity conservation of light quark and the angular momentum conservation. The $T_{4Q}^{1^{+-}}$ can decay into triple gluons or a gluon associated with a pair light quarks, which however is suppressed by the strong coupling constant $\alpha_s$. Thus,  we will not consider the decay of $T_{4Q}^{1^{+-}}$ in current work.}
\begin{subequations}\label{nrqcd-formula}
\begin{eqnarray}
\Gamma[T_{4Q}^{0^{++}}\to {\rm LH}]&=&c^{(0)}_{{\rm LH},1}\times \frac{2m_H\langle\mathcal{O}_{\mathbf{6}\otimes\mathbf{\bar 6}}^{(0)}\rangle}{4^2 (2m_Q)^4}
+c^{(0)}_{{\rm LH},2}\times \frac{2m_H\langle\mathcal{O}_{\mathbf{\bar 3}\otimes\mathbf{3}}^{(0)}\rangle}{4^2 (2m_Q)^4}
+c^{(0)}_{{\rm LH},3}\times \frac{2m_H\langle\mathcal{O}_{\rm mixing}^{(0)}\rangle}{4^2 (2m_Q)^4},\phantom{xxx}\\
\Gamma[T_{4Q}^{2^{++}}\to {\rm LH}]&=&c^{(2)}_{{\rm LH}}\times \frac{2m_H\langle\mathcal{O}_{\mathbf{\bar 3}\otimes\mathbf{3}}^{(2)}\rangle}{4^2 (2m_Q)^4},
\end{eqnarray}
\end{subequations}
where the superscript in $T_{4Q}$ denotes the quantum number $J^{PC}$ of the $T_{4Q}$, $m_Q$ and $m_H$ signify the masses of the heavy quark $Q$ and the $T_{4Q}$ respectively, and $c$ represent
the SDCs. 

The long-distance matrix elements (LDMEs) in (\ref{nrqcd-formula}) are defined via
\begin{subequations}\label{nrqcd-ldmes}
\begin{eqnarray}
\langle\mathcal{O}_{\mathbf{6}\otimes\mathbf{\bar 6}}^{(0)}\rangle &=&  \big|\langle 0| \mathcal{O}_{\mathbf{6}\otimes\mathbf{\bar 6}}^{(0)}|T_{4Q}\rangle\big|^2,\\
\langle\mathcal{O}_{\mathbf{\bar 3}\otimes\mathbf{3}}^{(J)}\rangle &=&  \big|\langle 0| \mathcal{O}_{\mathbf{\bar 3}\otimes\mathbf{3}}^{(J)}|T_{4Q}\rangle\big|^2,\\
\langle\mathcal{O}_{\rm mixing}^{(0)}\rangle &=& {\rm Re}\big[\langle 0| \mathcal{O}_{\mathbf{\bar 3}\otimes\mathbf{3}}^{(0)}|T_{4Q}\rangle \langle T_{4Q}| \mathcal{O}_{\mathbf{6}\otimes\mathbf{\bar 6}}^{(0)\dagger}|0\rangle \big],
\end{eqnarray}
\end{subequations}
where $J=0,  2$, and the $T_{4Q}$ state is nonrelativistically normalized. 
Note that, in Eq.~(\ref{nrqcd-ldmes}), we have made use of the ``vacuum-saturation approximation'' to transfer the hadronic matrix elements
into the electromagnetic matrix elements~\cite{Bodwin:1994jh}. 
The operators in (\ref{nrqcd-ldmes}) are constructed in the diquark-antidiquark basis~\footnote{Alternatively, the NRQCD operators can be constructed in $Q\bar{Q}$-$Q\bar{Q}$ basis, as done in Ref.~\cite{Zhu:2020xni},
where the $Q\bar{Q}$ cluster  can  be either in color singlet or color octet.  Actually, the NRQCD operators in Eq.~(\ref{nrqcd-operators}) can be converted to the operators in Ref.~\cite{Zhu:2020xni} by Fierz transformation.}, where the spin configuration and color configuration of the diquark are correlated due to Fermi statistics. Specifically, the S-wave spin-singlet diquark must be a color-sextet, while the S-wave spin-triplet diquark must be a color-triplet.
Explicitly, these operators read
\begin{subequations}\label{nrqcd-operators}
\begin{eqnarray}
\mathcal{O}_{\mathbf{6}\otimes\mathbf{\bar 6}}^{(0)}&=&  [\psi_a^T (i \sigma^2)\psi_b][\chi_c^\dagger (i \sigma^2)\chi_d^{*}]
{\mathcal C}^{ab;cd}_{\mathbf{6}\otimes\mathbf{\bar 6}},\\
\mathcal{O}_{\mathbf{\bar 3}\otimes\mathbf{3}}^{(0)}&=& -\frac{1}{\sqrt{3}} [\psi_a^T (i \sigma^2)\sigma^i\psi_b][\chi_c^\dagger \sigma^i (i \sigma^2)\chi_d^{*}]
{\mathcal C}^{ab;cd}_{\mathbf{\bar 3}\otimes\mathbf{3}},\\
\mathcal{O}_{\mathbf{\bar 3}\otimes\mathbf{3}}^{(2)}&=& \frac{1}{2}\Gamma^{ij,kl}\epsilon_H^{ij*} [\psi_a^T (i \sigma^2)\sigma^k\psi_b][\chi_c^\dagger \sigma^l (i \sigma^2)\chi_d^{*}]
{\mathcal C}^{ab;cd}_{\mathbf{\bar 3}\otimes\mathbf{3}},
\end{eqnarray}
\end{subequations}
where $\psi$ and $\chi^\dagger$ are Pauli spinor fields that
annihilate the heavy quark and antiquark, respectively, $\sigma^i$ denotes Pauli matrix, and $\epsilon_H$ denotes the polarization tensor of the $T_{4Q}^{2^{++}}$.
The Latin letters $i,j,k=1,2,3$ signify the Cartesian indices, whereas $a,b,c,d=1,2,3$ denote the color indices.
The symmetric traceless tensor is
\beq
\Gamma^{ij,kl}=\delta^{ik} \delta^{jl}+\delta^{il} \delta^{jk}-\frac{2}{3}\delta^{ij}\delta^{kl}.
\label{symetric:tensor}
\eeq
The color projection tensors in \eqref{nrqcd-operators} are given by
\begin{subequations}\label{color:tensor}
\begin{eqnarray}
\mathcal{C}^{ab;cd}_{\mathbf{6}\otimes\mathbf{\bar 6}}\equiv \frac{1}{2\sqrt{6}}(\delta^{ac}\delta^{bd}+\delta^{ad}\delta^{bc}),\\
\mathcal{C}^{ab;cd}_{\mathbf{\bar 3}\otimes\mathbf{3}}\equiv \frac{1}{2\sqrt{3}}(\delta^{ac}\delta^{bd}-\delta^{ad}\delta^{bc}).
\end{eqnarray}
\end{subequations}

It is straightforward to convert the factorization formula (\ref{nrqcd-formula}) into that for $T_{4Q}\to {\gamma\gamma}$.
Correspondingly, we replace the subscript ${\rm LH}$ with $\gamma\gamma$ in the SDCs to denote the SDCs for $T_{4Q}\to {\gamma\gamma}$.

\subsection{Procedure to determine the SDCs}
 With the spirit of factorization, the SDCs in (\ref{nrqcd-formula}) are insensitive to the hadronization effects 
 in the tetraquark, thus they can be deduced with the aid of the standard perturbative matching techniques.
 That is, by replacing the physical $T_{4Q}$ state in (\ref{nrqcd-formula}) with a fictitious ``tetraquark" state composed of a pair of heavy quarks and a pair of heavy antiquarks, carrying the same quantum number as the physical $T_{4Q}$.
Conveniently, we label the fictitious state with $\widetilde{T}_{4Q}$.
 After this replacement,
we can compute both sides of (\ref{nrqcd-formula}) order by order in $\alpha_s$, thus the SDCs can be solved at desired order of $\alpha_s$.

To calculate the left-hand side of (\ref{nrqcd-formula}),  we first write down the amplitude of the free $QQ\bar{Q}\bar{Q}$ decay, then enforce the 
$QQ\bar{Q}\bar{Q}$
in the desired spin, total angular momentum, and color quantum numbers.
In a shortcut, we employ the spin-singlet projector $\Pi_0$ and spin-triplet projector $\Pi_1$ to enforce the diquark in $S=0$ and $S=1$ respectively, 
and we make use of the color projectors  $\mathcal{C}^{ab;cd}_{\mathbf{6}\otimes\mathbf{\bar 6}}$ and 
$\mathcal{C}^{ab;cd}_{\mathbf{\bar 3}\otimes\mathbf{3}}$ to extract the color-sextet and color-triplet parts of the diquark, respectively. 
For the case of the diquark and anti-diquark in the spin-triplet configuration, we apply the covariant projectors to pick out the total angular momentum number of $\widetilde{T}_{4Q}$.  
Concretely, we ensure the $QQ\bar{Q}\bar{Q}$ in $J^{PC}=0^{++}$ through the replacement
\begin{eqnarray}\label{eq-replacement-66bar}
u^a_i u^b_j \bar{v}^c_k \bar{v}^d_l \to (C\Pi_0)^{ij} (\Pi_0 C)^{lk} \mathcal{C}^{ab;cd}_{\mathbf{6}\otimes\mathbf{\bar 6}},
\end{eqnarray}
for the spin-singlet diquark configuration,
and ensure the $QQ\bar{Q}\bar{Q}$ in $J^{PC}=0^{++},  2^{++}$ through the replacement
\begin{eqnarray}\label{eq-replacement-33bar}
u^a_i u^b_j \bar{v}^c_k \bar{v}^d_l \to (C\Pi_{1\mu})^{ij} (\Pi_{1\nu} C)^{lk} \mathcal{C}^{ab;cd}_{\mathbf{\bar 3}\otimes\mathbf{3}} J^{\mu\nu}_{0,2},
\end{eqnarray}
for the spin-triplet diquark configuration, where $u$ and $v$ denote the Dirac spinors of heavy quarks.
In (\ref{eq-replacement-66bar}) and (\ref{eq-replacement-33bar}), the spin projectors are
\begin{subequations}\label{spin-projector}
 \begin{eqnarray}
\Pi_0=\frac{\left({P\!\!\!\slash\over 4}- m_Q\right)\gamma_5}{\sqrt{2}},\\
\Pi_1^\mu=\frac{\left({P\!\!\!\slash\over 4}- m_Q\right)\gamma^\mu}{\sqrt{2}},
\end{eqnarray}
\end{subequations}
 and the covariant projectors $J^{\mu\nu}_{0, 2}$ are
\begin{subequations}\label{spin-projector-1-1}
 \begin{eqnarray}
J^{\mu\nu}_0&=&\frac{1}{\sqrt{3}}\eta^{\mu\nu},\\
J^{\mu\nu}_2&=&\epsilon_{H,\alpha\beta}[\frac{1}{2}\eta^{\mu\alpha}\eta^{\nu\beta}+\frac{1}{2}\eta^{\mu\beta}\eta^{\nu\alpha}
-\frac{1}{3}\eta^{\mu\nu}\eta^{\alpha\beta}],
\end{eqnarray}
\end{subequations}
where $P$ denotes the momentum of the $T_{4Q}$, and $\eta^{\mu\nu}\equiv -g^{\mu\nu}+\tfrac{P^\mu P^\nu}{P^2}$.

Furthermore, the squared amplitude can be obtained by multiplying the amplitude with its complex conjugate, 
summing the polarizations of the final states, and averaging the polarizations of the $\widetilde{T}_{4Q}$.
It is worth noting that, in order to match $c_{\gamma\gamma,3}^{(0)}$ and $c_{\rm LH,3}^{(0)}$, the SDCs of the mixing LDME $\langle\mathcal{O}_{\rm mixing}^{(0)}\rangle$, one must utilize the replacement (\ref{eq-replacement-33bar}) for the quark-level amplitude, and the replacement (\ref{eq-replacement-66bar}) for the complex conjugate of the amplitude.

For the sake of completeness, we present the results for the perturbative LDMEs that involve $\widetilde{T}_{4Q}$
 \begin{eqnarray}\label{ldme-perturbative}
\big|\langle 0| \mathcal{O}_{\mathbf{6}\otimes\mathbf{\bar 6}}^{(0)}|\tilde{T}_{4Q}\rangle\big|^2=\big|\langle 0| \mathcal{O}_{\mathbf{\bar 3}\otimes\mathbf{3}}^{(J)}|\widetilde{T}_{4Q}\rangle\big|^2={\rm Re}\big[\langle 0| \mathcal{O}_{\mathbf{\bar 3}\otimes\mathbf{3}}^{(0)}|\widetilde{T}_{4Q}\rangle \langle \widetilde{T}_{4Q}| \mathcal{O}_{\mathbf{6}\otimes\mathbf{\bar 6}}^{(0)\dagger}|0\rangle \big]=4^2(2m_Q)^4,
\end{eqnarray}
where the heavy quark states are relativistically normalized, which is consistent with the convention adopted for the spin projectors (\ref{spin-projector}).

Now, we collect all the essential ingredients to evaluate the SDCs.

\section{Analytic expressions for the SDCs~\label{sec-sdcs}}

\begin{figure}[htbp]
	\centering
	\includegraphics[width=0.95\textwidth]{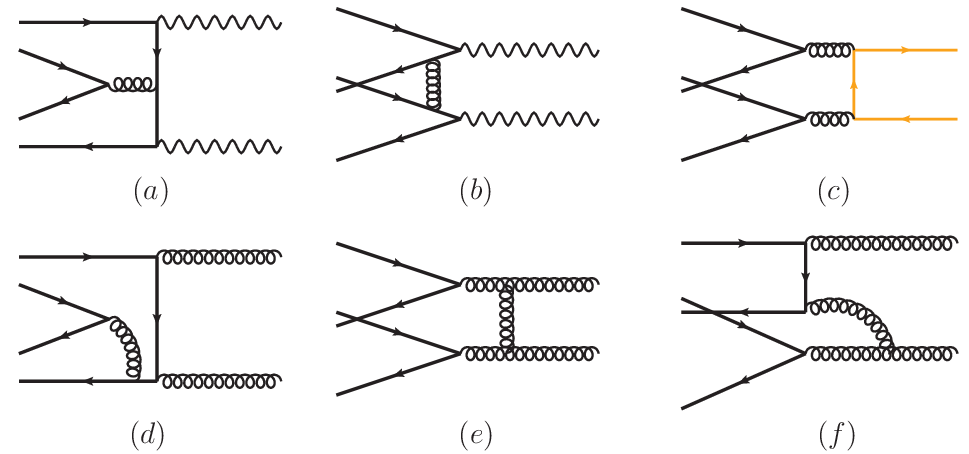}
	\caption{Some representative Feynman diagrams for $T_{4Q}\to\gamma\gamma$, $T_{4Q}\to q\bar{q}$, and $T_{4Q}\to gg$.
	\label{fig-feynman-diagram-photons}}
\end{figure}

We use the package FeynArts~\cite{Hahn:2000kx} to generate quark-level Feynman diagrams and the corresponding amplitudes.
There are 40, 4, and 62 nonvanishing Feynman diagrams for $T_{4Q}\to \gamma\gamma$, $T_{4Q}\to q\bar{q}$, and $T_{4Q}\to  gg$, respectively. 
Some representative diagrams are illustrated in Fig.~\ref{fig-feynman-diagram-photons}.
After implementing the replacements (\ref{eq-replacement-66bar}) and (\ref{eq-replacement-33bar}) to ensure $QQ\bar{Q}\bar{Q}$ in the correct quantum numbers, we use the packages FeynCalc\cite{Mertig:1990an} and FormLink\cite{Feng:2012tk,Kuipers:2012rf} 
to conduct the trace over Dirac and $SU(N_c)$ color matrices, and the contraction over Lorentz indices.

Following the matching procedure sketched in Sec.~\ref{sec-framework}, it is straightforward to deduce the SDCs\\
\begin{subequations}\label{sdcs-expand}
	\begin{eqnarray}
	c^{(0)}_{\gamma\gamma,1}&=&\frac{32 \pi ^3 e_Q^4 \alpha^2 \alpha_s^2}{3 m_H m_Q^4},\\
	c^{(0)}_{\gamma\gamma,2}&=&\frac{576 \pi ^3 e_Q^4 \alpha^2 \alpha_s^2}{ m_H m_Q^4},\\
	c^{(0)}_{\gamma\gamma,3}&=&\frac{64\sqrt 6 \pi ^3 e_Q^4 \alpha^2 \alpha_s^2}{m_H m_Q^4},\\
	c^{(2)}_{\gamma\gamma}&=&\frac{4096 \pi ^3 e_Q^4 \alpha^2 \alpha_s^2}{15 m_H m_Q^4},
	\end{eqnarray}
\end{subequations}
for $T_{4Q} \to \gamma\gamma$, and
\begin{subequations}\label{sdcs-expand}
	\begin{eqnarray}
	c^{(0)}_{{\rm LH},1}&=&\frac{121 \pi^3 \alpha_s^4}{27m_H m_Q^4},\\
	c^{(0)}_{{\rm LH},2}&=&\frac{18 \pi^3 \alpha_s^4}{m_H m_Q^4},\\
	c^{(0)}_{{\rm LH},3}&=&\frac{22\sqrt 6 \pi^3 \alpha_s^4}{3m_H m_Q^4},\\
	c^{(2)}_{{\rm LH}}&=&\frac{1568 \pi^3 \alpha_s^4}{135 m_H m_Q^4} \left(1+\frac{48}{49}n_L\right),
	\end{eqnarray}
\end{subequations}
for $T_{4Q} \to {\rm LH}$, where $n_L=3$ represents the number of light quark flavors.  Note that 
the $n_L$ term in $c^{(2)}_{{\rm LH}}$ corresponds to the contribution from $T_{4Q}\to q\bar{q}$.
It is worth emphasizing that the $m_H$ in the SDCs,  originating from the prefactor of the formula of decay width, cancels the same factor  in (\ref{nrqcd-formula}), thus the final decay widths are free of the $m_H$.    

\section{Phenomenological predictions~\label{sec-phen}}

Prior to making phenomenological predictions, we need to fix the various input parameters.~\footnote{Since exact values of the LDMEs of $T_{4b}$ are currently absent in the literature,  in this section, we are mainly concerned with predictions for $T_{4c}$, and briefly estimate the decay widths for $T_{4b}$.}
We take the charm quark mass to be $m_c=1.5$ GeV. 
We take the fine structure coupling constant $\alpha=1/137$. The QCD running coupling constant at $\mu_R=2 m_c$ is evaluated with the aid of the package RunDec~\cite{Chetyrkin:2000yt}.  
We have varied $\mu_R$ from $m_c$ to $4m_c$ to estimate the theoretical uncertainty.

We should further choose the NRQCD LDMEs. 
These nonperturbative LDMEs can be related to the phenomenological four-body Schr\"odinger wave functions at the origin~\cite{Feng:2023agq}
\begin{subequations}
\begin{align}
 &\langle\mathcal{O}_{\mathbf{\bar 3}\otimes\mathbf{3}}^{(J)}\rangle \approx 16\psi_{3}^{(J)}(0)\psi_{3}^{(J)*}(0), \\
&\langle\mathcal{O}_{\mathbf{6}\otimes\mathbf{\bar 6}}^{(0)}\rangle\approx 16\psi_{6}(0)\psi_{6}^*(0),\\
&\langle\mathcal{O}_{\rm mixing}^{(0)}\rangle\approx 16 |\psi_{3}^{(0)}(0)\psi_{6}^*(0)|,
\end{align}
\label{NRQCD:composite:operators}
\end{subequations}
In this work, we adopt two phenomenological models to evaluate the LDMEs~\cite{Lu:2020cns,liu:2020eha}. 
In both models, Cornell-type potentials with spin-dependent corrections are assumed, and  Gaussian basis method is utilized to solve for the four-body tetraquark
wave functions. However unlike the Model I~\cite{Lu:2020cns}, which is based on nonrelativistic quark model,  there involves a relativistic kinetic term in Model II~\cite{liu:2020eha}.
We enumerate the values of the LDMEs for $T_{4c}$ in Table~\ref{tab:LDME}. Note that there is a sign difference for the value of 
$\langle\mathcal{O}_{\rm mixing}^{(0)}\rangle$ from the two models. 
\begin{table}[!htbp]\small
	\caption{ Numerical values of the LDMEs for $T_{4c}$ in Model I~\cite{Lu:2020cns} and Model II~\cite{liu:2020eha}, in unit of $\rm GeV^{9}$. }
	\label{tab:LDME}
	\centering
	\setlength{\tabcolsep}{20pt}
	\renewcommand{\arraystretch}{1.6}
	{
		\begin{tabular}{|c|c|c|c|c|c|c|}
			\hline
			$\rm J^{PC}$
			&\multicolumn{3}{|c|}{$0^{++}$}
			&$1^{+-}$
			&$2^{++}$\\
			\hline
			LDMEs
			&$\langle\mathcal{O}_{\mathbf{\bar 3}\otimes\mathbf{3}}^{(0)}\rangle$
			&$\langle\mathcal{O}_{\rm mixing}^{(0)}\rangle$
			&$\langle\mathcal{O}_{\mathbf{6}\otimes\mathbf{\bar 6}}^{(0)}\rangle$
			&$\langle\mathcal{O}_{\mathbf{\bar 3}\otimes\mathbf{3}}^{(1)}\rangle$
			&$\langle\mathcal{O}_{\mathbf{\bar 3}\otimes\mathbf{3}}^{(2)}\rangle$\\
			\hline
			Model I
			&$0.0347$
			&$0.0211$
			&$0.0128$
			&$0.0260$
			&$0.0144$\\
			\hline
			Model II
			&$0.0187$
			&$-0.0161$
			&$0.0139$
			&$0.0160$
			&$0.0126$\\
			\hline
		\end{tabular}
	}
\end{table}

If assuming $T_{4c}$ decay is saturated by its decay into double $J/\psi$, we can approximate 
the total decay width of $T_{4c}$ through
\begin{eqnarray}\label{eq-total-decay-rate}
\Gamma_{\rm total}(T_{4c})\approx \Gamma[T_{4c}\to J/\psi J/\psi]\approx 0.12\, {\rm GeV},
\end{eqnarray}
where $0.12$ GeV corresponds to the central value of the decay width determined by the ATLAS collaboration~\cite{ATLAS:new1},  and is
roughly the average of the two fitting values from the LHCb measurement~\cite{LHCb:2020bwg}. 
This value is also consistent with the no-interference fitting value by the  CMS collaboration~\cite{CMS:2023owd}.

Now, we collect all ingredients to make phenomenological predictions. 
The theoretical results of the decay widths as well as the corresponding branching fractions for $T_{4c}\to \gamma\gamma$ and $T_{4c}\to {\rm LH}$ are tabulated in Table~\ref{tab-2photons}. 
We observe that,  the branching fractions for $T_{4c}\to {\rm LH}$ are about three order-of-magnitude larger than these for $T_{4c}\to \gamma\gamma$,
which is attributed to enhancement from the strong coupling constant, i.e., $\alpha_s^2/\alpha^2\approx 10^3$. 
By comparing the predictions from the two phenomenological models,  we find the theoretical results for $2^{++}$ tetraquark are insensitive to the models,
while the predictions for $0^{++}$ from Model I are more than two times larger than from Model II.  
As can be seen from Table I, the values of $\langle\mathcal{O}_{\rm mixing}^{(0)}\rangle$ in the two models  take different signs. Therefore,
the interfering term is constructive in Model I, while destructive in Model II, which explains why the branching fractions  for $0^{++}$ from Model I are larger.
Finally, it is worth noting that our prediction for the ${\rm Br}[T_{4c}^{0^{++}}\to \gamma\gamma]$ from Model I is slightly smaller than the value $(2.77\pm0.36)\times 10^{-6}$ estimated 
based on the vector meson dominance~\cite{Biloshytskyi:2022pdl}.

\begin{table}[!htbp]\small
	\caption{Theoretical predictions on the decay widths and branching fractions.
		We estimate the uncertainty by sliding $\mu_R$ from $m_c$ to $4m_c$. }
	\label{tab-2photons}
	\centering
	\setlength{\tabcolsep}{10pt}
	\renewcommand{\arraystretch}{1.6}
	{
		\begin{tabular}{|c|c|c|c|c|}
			\hline
			\multicolumn{5}{|l|}{$m_c=\rm 1.5 GeV$; $\Gamma(\times10^{-1}\rm MeV) $ }\\			\hline
			\multirow{2}{*}{$\rm channel$}
			&\multicolumn{2}{|c|}{Model I}
			&\multicolumn{2}{|c|}{Model II}
			\\
			\cline{2-5}
			&$\Gamma_{\rm LO}$
			&Br
			&$\Gamma_{\rm LO}$
			&Br
			\\
			\hline
			\multirow{1}{*}{$T_{4c}^{0^{++}}\to \gamma\gamma$}
			& $1.29^{+0.82}_{-0.41}\times10^{-3}$
			& $1.08^{+0.68}_{-0.34}\times10^{-6}$
			& $4.63^{+2.94}_{-1.47}\times10^{-4}$
			& $3.86^{+2.45}_{-1.23}\times10^{-7}$
			\\
			\hline
			\multirow{1}{*}{$T_{4c}^{2^{++}}\to \gamma\gamma$}
			& $0.22^{+0.14}_{-0.07}\times10^{-3}$
			& $0.18^{+0.11}_{-0.06}\times10^{-6}$
			& $1.90^{+1.21}_{-0.60}\times10^{-4}$
			& $1.58^{+1.01}_{-0.50}\times10^{-7}$
			\\
			\hline
			\multirow{1}{*}{$T_{4c}^{0^{++}}\to\rm LH$}
			& $0.31^{+0.52}_{-0.16}$
			& $0.26^{+0.43}_{-0.14}\times 10^{-3}$
			& $0.03^{+0.05}_{-0.02}$
			& $0.27^{+0.45}_{-0.14}\times 10^{-4}$
			\\
			\hline
			\multirow{1}{*}{$T_{4c}^{2^{++}}\to \rm LH$}
			& $0.19^{+0.32}_{-0.10}$
			& $0.16^{+0.27}_{-0.09}\times 10^{-3}$
			& $0.17^{+0.28}_{-0.09}$
			& $1.40^{+2.34}_{-0.75}\times 10^{-4}$
			\\
			\hline
			
		\end{tabular}
	}
\end{table}

To evaluate the numerical values for $T_{4b}$ decay, we should determine the LDMEs. The exact values of the LDMEs for $T_{4b}$ are absent in the literature. 
As a very crude estimate~\cite{Feng:2023agq}, we temporarily assume
that $T_{4Q}$ is comprised of a compact diquark-antidiquark cluster, each of which is bound by attractive color Coulomb forces. 
One then estimates the ratio of the four-body
Schr\"odinger wave functions at the origin for $T_{4c}$ and for $T_{4b}$ through simple dimensional analysis. Hence, we may roughly estimate the LDMEs of the $T_{4b}$ by
\begin{eqnarray}
\langle O^{(J)}\rangle_{T_{4b}}=\langle O^{(J)}\rangle_{T_{4c}}\times \frac{\langle O\rangle_{T_{4b},\rm Coulomb}}{\langle O\rangle_{T_{4c},\rm Coulomb}}
\approx\langle O^{(J)}\rangle_{T_{4c}}\times\bigg(\frac{m_b\alpha_s^b}{m_c\alpha_s^c}\bigg)^9,
\end{eqnarray}
where $\alpha_s^Q$
represents the strong coupling $\alpha_s(m_Q v_Q)\sim v_Q$,
$v_Q$ stands for the typical velocity of the heavy quark inside the tetraquark, and the subscript ``Coulomb''
indicates that the LDME is evaluated using the
diquark model with the interquark and interdiquark potentials being Coulombic. By taking $m_b=4.8$ GeV, $v_b=\sqrt{0.1}$, $v_c=\sqrt{0.3}$,
and $\alpha_s(2m_b)=0.175$,
we predict $\Gamma[T_{4b}^{0^{++}}\to \gamma\gamma]=(0.4-1.0)\times 10^{-7}$ MeV,  
$\Gamma[T_{4b}^{2^{++}}\to \gamma\gamma]=(1.5-1.8)\times 10^{-8}$ MeV, $\Gamma[T_{4b}^{0^{++}}\to \rm LH]=(0.2-2.2)\times 10^{-4}$ MeV, and $\Gamma[T_{4b}^{2^{++}}\to \rm LH]=(1.2-1.4)\times 10^{-4}$ MeV. We find the decay width of $T_{4b}\to \gamma\gamma$ is roughly three order-of-magnitude smaller than that of $T_{4c}$, while the decay width of $T_{4b}\to \rm LH$ is two order-of-magnitude smaller than the $T_{4c}$ counterpart.

Combing our predictions with the $T_{4c}$ production cross sections at the LHC~\cite{Feng:2023agq}, we  can further estimate the event yields for the $T_{4c}$ 
hadronic and electromagnetic decay at the LHC. The event numbers for various channels are tabulated in Table~\ref {tab-event-number}.
For reference, we also copy the $T_{4c}$ cross sections from Ref.~\cite{Feng:2023agq} in the table.
It is expected that there will be plenty of $T_{4c}\to {\rm LH}$ signals accumulated at the LHC. 
In spite of potentially copious background events,  it seems that the observation prospects for the $T_{4c}$ hadronic decay are promising.
On the other hand, the event numbers for the electromagnetic decay $T_{4c}\to \gamma\gamma$ are  much smaller, 
nevertheless,  it is hopeful to probe these channels,  thanks to a clean final state topology in which the diphoton invariant mass can be reconstructed with high precision.

\begin{table}[!htbp]\small
	\caption{Estimation on the event yields at the LHC, where the $T_{4c}$ production cross sections at the LHC $\sigma_{\rm LHC}$ are taken from Ref.~\cite{Feng:2023agq} and the integrated luminosity $\mathcal{L}=100\, {\rm fb}^{-1}$ is chosen. 
		The two uncertainties in $N_{\rm events}$ are transferred from the uncertainties in the cross sections and branching fractions, respectively.}
	\label{tab-event-number}
	\centering
	\setlength{\tabcolsep}{10pt}
	\renewcommand{\arraystretch}{1.6}
	{
		\begin{tabular}{|c|c|c|c|c|}
			\hline
			
			\multirow{2}{*}{$\rm channel$}
			&\multicolumn{2}{|c|}{Model I}
			&\multicolumn{2}{|c|}{Model II}
			\\
			\cline{2-5}
			&$\sigma_{\rm LHC}[\rm nb]$
			&$N_{\rm events}$
			&$\sigma_{\rm LHC}[\rm nb]$
			&$N_{\rm events}$
			\\
			\hline
			\multirow{1}{*}{$T_{4c}^{0^{++}}\to \gamma\gamma$}
			& $37\pm 26$
			& $(3.98 \pm 2.80 ^{+2.53}_{-1.28})\times 10^3$
			& $9 \pm 6$
			& $(3.47 \pm 2.32 ^{+2.21}_{-1.10})\times 10^2$
			\\
			\hline
			\multirow{1}{*}{$T_{4c}^{2^{++}}\to \gamma\gamma$}
			& $93 \pm 65$
			& $(1.68 \pm 1.18 ^{+1.07}_{-0.53})\times 10^3$
			& $81 \pm 57$
			& $(1.28 \pm 0.90^{+0.81}_{-0.41})\times 10^3$
			\\
			\hline
			\multirow{1}{*}{$T_{4c}^{0^{++}}\to\rm LH$}
			& $37\pm 26$
			& $(0.95 \pm 0.67 ^{+1.59}_{-0.51})\times 10^6$
			& $9 \pm 6$
			& $(2.40 \pm 1.60 ^{+4.01}_{-1.28})\times 10^4$
			\\
			\hline
			\multirow{1}{*}{$T_{4c}^{2^{++}}\to \rm LH$}
			& $93 \pm 65$
			& $(1.49 \pm 1.04^{+2.49}_{-0.79})\times 10^6$
			& $81 \pm 57$
			& $(1.13 \pm 0.80^{+1.89}_{-0.60})\times 10^6$
			\\
			\hline
			
		\end{tabular}
	}
\end{table}
\section{Summary~\label{sec-sum}}
By applying the NRQCD factorization formalism, we compute the decay widths for  
the S-wave fully heavy tetraquark $T_{4Q}$ hadronic as well as electromagnetic decay 
at lowest order in $\alpha_s$ and $v$. 
The SDCs are determined through the procedure of perturbative matching. 
The LDMEs are related to the phenomenological four-body Schr\"odinger wave functions at the origin, 
whose values are taken from Refs.~\cite{Lu:2020cns,liu:2020eha}. 
To obtain the branching fractions for $T_{4c}$ decay,
we approximate the total decay width of $T_{4c}$ with $\Gamma[T_{4c}\to J/\psi J/\psi]$. The latter value has
been determined by the LHCb, CMS, and ATLAS collaborations. 
We find the branching fractions are around $10^{-4}$ and $10^{-7}-10^{-6}$ for the $T_{4c}$ hadronic decay and electromagnetic decay, respectively.
It is worth noting that the branching fractions for $T_{4c}^{2^{++}}$ are insensitive to the phenomenological models, 
while the predictions for $T_{4c}^{0^{++}}$ from Ref.~\cite{Lu:2020cns} are more than two times larger than these from Ref.~\cite{liu:2020eha}.  
This feature is quite similar to the case for the $T_{4c}$ production at the LHC. 
Combing the $T_{4c}$ production cross sections with the branching fractions for $T_{4c}$ decay, we estimate the event numbers for various decay channels
at the LHC. The observation prospect seems to be promising for both the $T_{4c}$ hadronic and electromagnetic decay. 

\begin{acknowledgments}
The work of W.-L. S. and T. W. is supported by the National Natural Science Foundation of China under Grants No. 12375079 and No. 11975187, and the Natural Science
Foundation of ChongQing under Grant No. CSTB2023NSCQ-MSX0132.
The work of Y.-D. Z. is supported by the National Natural Science Foundation
of China under Grant No.~12135006 and No.~12075097, as well as by the Fundamental Research Funds for the Central Universities under Grant No.~CCNU20TS007 and No.~CCNU22LJ004.
 The work of F. F. is supported by the National Natural Science Foundation of China under Grant No.~12275353 and No.~11875318.
\end{acknowledgments}


\end{document}